\documentclass{article}
\usepackage{spconf,amsmath,graphicx}
\usepackage{amsmath,amsfonts,bm}
\usepackage{hyperref}
\usepackage{url}
\usepackage{booktabs}
\usepackage{multirow}
\usepackage{makecell}

\DeclareMathOperator*{\argmax}{arg\,max}

\title{FoolHD: Fooling speaker identification by Highly imperceptible adversarial Disturbances}

\name{\begin{tabular}{c}Ali Shahin Shamsabadi$^{\star}$~$^1$, Francisco Sepúlveda Teixeira$^{\star}$~$^2$\thanks{$^{\star}$Authors contributed equally to this work.}, Alberto Abad~$^2$, Bhiksha Raj~$^3$,\\
Andrea Cavallaro~$^1$, Isabel Trancoso~$^2$\end{tabular}}
\address{$^1$Centre for Intelligent Sensing - Queen Mary University of London,\\ $^2$INESC-ID/IST - University of Lisbon, $^3$Carnegie Mellon University
\thanks{The authors would like to thank Thomas Rolland and Catarina Botelho for their contributions in the implementation of the \textit{x-vector} speaker identification network. This work was supported by Portuguese national funds through Fundação para a Ciência e a Tecnologia (FCT), with references UIDB/
 50021/2020 and  CMU/TIC/0069/2019, and also BD2018 ULisboa. Bhiksha Raj was supported by the U.S. Army Research Laboratory and DARPA under contract HR001120C0012. We also wish to thank the Alan Turing Institute (EP/N510129/1), which is funded by the U.K. Engineering and Physical Sciences Research Council, for its support through the project PRIMULA.}}

\begin{document}
\ninept   
\maketitle

\begin{abstract}
Speaker identification models are vulnerable to carefully designed adversarial perturbations of their input signals that induce misclassification.
In this work, we propose a white-box steganography-inspired adversarial attack that generates imperceptible adversarial  perturbations against a speaker identification model.
Our approach, FoolHD, uses a Gated Convolutional Autoencoder that operates in the DCT domain and is trained with a multi-objective loss function, to generate and conceal the adversarial perturbation within the original audio files. In addition to hindering speaker identification performance, this multi-objective loss accounts for human perception through a frame-wise cosine similarity between MFCC feature vectors extracted from the original and adversarial audio files. We validate the effectiveness of FoolHD with a 250-speaker identification x-vector network, trained using VoxCeleb, in terms of accuracy, success rate, and imperceptibility.
Our results show that FoolHD generates highly imperceptible adversarial audio files (average PESQ scores above $4.30$), while achieving a success rate of 99.6\% and 99.2\% in misleading the speaker identification model, for untargeted and targeted settings, respectively.
\end{abstract}
\vspace{-0.5cm}
\section{Introduction}
\label{sec:intro}

Recent years have seen a significant increase in the availability and use of deep learning-based speech applications~\cite{mclean2019}. The ubiquitous nature of speech makes such applications useful and easily accessible. Automatic speaker identification is a core technology for many of these applications, which relies heavily on deep learning models~\cite{nassif2019}. However, these models have been shown to be vulnerable to adversarial attacks -- i.e. the perturbation of a classifier's input at test time, such that the classifier outputs a wrong prediction~\cite{szegedy2013intriguing}.
Adversarial attacks can be untargeted or targeted~\cite{abdullah2020faults}.
An \textit{untargeted} attack pushes the model to misidentify the speaker for a given audio file, whereas a \textit{targeted} attack attempts to force the model to identify a specific speaker, which is chosen by the attacker. The knowledge of the attacker about the speaker identification model can also vary~\cite{abdullah2020faults}. In a white-box setting, all the model-related information, such as its architecture and parameters, is available to the attacker (e.g.~open-source models). Contrarily, in a black-box setting, the attacker's knowledge consists at most of the model's outputs (e.g.~public APIs). Adversarial perturbations should be imperceptible so that humans are not able to perceive the adversarial perturbation, nor distinguish the adversarial audio files from their corresponding original audio files~\cite{abdullah2020faults}.

Most of the existing adversarial attacks~\cite{jati2020adversarial} against speaker identification models exploit state-of-the-art methods originally developed for image classification, such as the Fast Gradient Sign Method (FGSM)~\cite{goodfellow2014} and its iterative version, Basic Iterative Method (BIM)~\cite{kurakin2016adversarial}.
Kreuk et al.~\cite{kreuk2018fooling} and Li et al.~\cite{liX2020adv} explored the vulnerabilities of \textit{x-vector} and \textit{i-vector} based speaker verification models to FGSM adversarial attacks. 
Li et al.~\cite{liZ2020practical} further integrated an estimate of room impulse responses with FGSM to generate adversarial audio files that may still be effective when played over-the-air against an \textit{x-vector} based speaker recognition model. 
Li et al.~\cite{liJ2020universal} tried to learn universal adversarial perturbations by adversarially training a perturbation generator against a SincNet-based speaker identification model. Although the aforementioned attacks achieve high success rates in misleading classifiers, most of them present high levels of distortion as they neglect the impact that adversarial perturbations have in human perception. For example, the PESQ~\cite{hu2007evaluation} score of the adversarial audio files generated by the universal adversarial perturbation of~\cite{liJ2020universal} is only 3 (out of 5). Recently, Wang et al.~\cite{wang2020inaudible} improved the imperceptibility of an FGSM-based attack against an \textit{x-vector} speaker identification by exploiting frequency masking. However, their method trades off imperceptibility against the success rate of their adversarial audio files (i.e. at the highest imperceptibility score of $4.23$, as measured by PESQ~\cite{hu2007evaluation}, the success rate is of $\sim73\%$).

In this work, we propose a novel adversarial attack to generate imperceptible adversarial audio perturbations by leveraging steganography techniques. As stated by Ian Goodfellow, the creation of adversarial examples can be seen as ``\textit{accidental steganography}''~\cite{goodfellow2014}. Even though adversarial examples and steganography have different goals, both try to hide a message in a carrier, such that this message does not perceptually affect the carrier. 
To this end, our proposed attack, FoolHD, adapts the speech steganography method proposed by~\cite{kreuk2019hide}, where a frequency-domain Gated Convolutional Autoencoder~\cite{dauphin2017language} is exploited to embed one or more audio files (i.e. messages) in another audio file (i.e. carrier). In particular, we train the Gated Convolutional Autoencoder to generate adversarial audio files whose perturbations are imperceptible to the human auditory system, against a white-box speaker identification model. We achieve these contrasting objectives with a novel multi-objective loss function that combines a perceptual loss function and an adversarial loss function. The former tries to make the adversarial audio files perceptually close to the original audio files, while the latter aims to mislead the speaker identification model in both \textit{untargeted} and \textit{targeted} settings.

\section{FoolHD}

Let $\mathbf{x} \in \mathbb{R}^{1\times D}$ be an (original) audio file and $f(\cdot)$ be an $N$-speaker identification model that predicts the most likely speaker
\begin{equation}
    y = f(\mathbf{x}) = \argmax_{i=1,...,N} p_i,
\end{equation}
where $p_i$ is the predicted probability of speaker $i$, computed by normalizing the $i$-th predicted logits, $z_i$, using a softmax operation:
\begin{equation}
    p_i=\frac{e^{z_i}}{\sum_{n=1}^Ne^{z_n}}.
\end{equation}
We aim to generate an adversarial audio file, $\dot{\mathbf{x}} \in \mathbb{R}^{1\times D}$, by perturbing $\mathbf{x}$ such that it \textit{misleads} the speaker identification model 
while ensuring this audio file remains perceptually similar to the original audio file (\textit{imperceptible perturbation}).

\begin{figure}[t!]
    \centering
    \includegraphics[width=0.95\columnwidth]{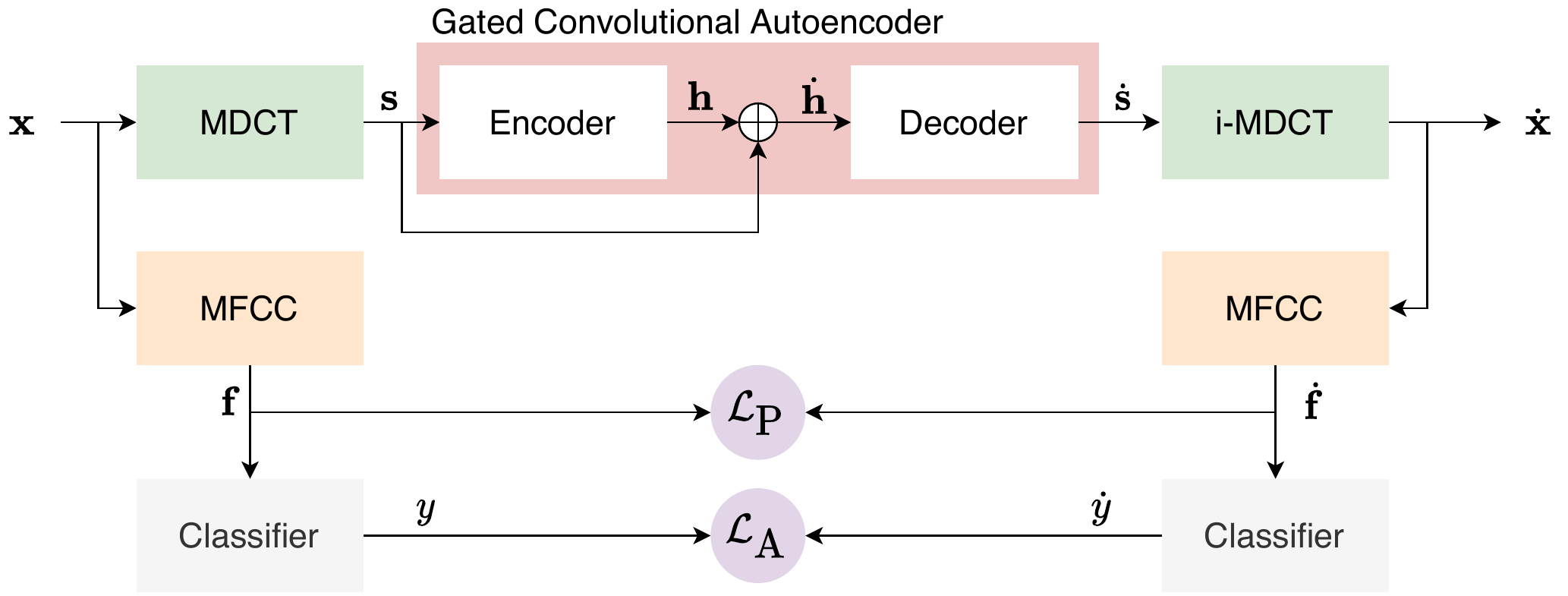}
    \caption{Overview of our proposed attack, FoolHD. 
    We obtain frequency-domain representations $\mathbf{s}$ of an audio file $\mathbf{x}$ via the Modified Discrete Cosine Transform (MDCT) and feed it to the encoder to produce the encoded representation of the spectrogram $\mathbf{h}$. The decoder generates the adversarial spectrogram $\dot{\mathbf{s}}$ by receiving $\dot{\mathbf{h}}=[\mathbf{s};\mathbf{h}]$ that is the concatenation of the original spectrogram and encoded original spectrogram, $\mathbf{h}$. A perceptual loss $\mathcal{L}_{\text{P}}(\cdot,\cdot)$ measures the perceptual similarity between the original audio and adversarial audio $\dot{\mathbf{x}}$ files. The perceptual loss is computed over the original, $\mathbf{f}$, and adversarial features, $\dot{\mathbf{f}}$. In addition to this, an adversarial loss $\mathcal{L}_{\text{A}}(\cdot,\cdot)$ quantifies the differences of the speaker identification model over the original and adversarial audio files.}
    
    \label{fig:BlockDiagram}
\end{figure}

\subsection{Overview} 

We propose to generate imperceptible adversarial perturbations by training a Gated Convolutional Autoencoder (GCA) composed of an encoder and a decoder, operating in the frequency domain (see Figure~\ref{fig:BlockDiagram}).
The architecture of our GCA is inspired by the steganographic method of~\cite{kreuk2019hide} in the sense of the encoder and decoder. The encoder, $\text{E}(\cdot)$, creates a latent representation of the spectral representation $\mathbf{s}$ of the original audio file, $\mathbf{h}=\text{E}(\mathbf{s})$ through three gated convolutional layers.
This representation $\mathbf{h}$ is then concatenated with the original input using a skip connection, to obtain the joint representation as $\dot{\mathbf{h}}=[\mathbf{h};\mathbf{s}]$ where $;$ denotes the concatenation operation. The decoder, $\mathbf{D}(\cdot)$, takes $\dot{\mathbf{h}}$ and generates the adversarial spectral representation of the audio as $\dot{\mathbf{s}}=\text{D}(\dot{\mathbf{h}})$ through four gated convolutional layers. Each gated convolutional layer of encoder and decoder, similarly to~\cite{kreuk2019hide}, is composed of 64 $3 \times 3$ kernels, followed by a batch normalization and a dropout layer.

However, our GCA has two main differences from~\cite{kreuk2019hide}. 
First, our GCA operates in a variation of the DCT type IV, called the Modified Discrete Cosine Transform (MDCT)~\cite{mdct}, as opposed to the Short-time Fourier Transform (STFT) domain. The STFT is a complex-valued transformation that represents an audio file with its magnitude and phase. Therefore, the phase and magnitude of the signal are processed separately resulting in reconstruction errors during the STFT inversion~\cite{kreuk2019hide}. 
This reconstruction error can be avoided by using the MDCT, as the MDCT is a real-valued transform, whose frequency coefficients encode both the phase and magnitude of the signal~\cite{zhang2013mdct}. Second, instead of providing an external steganographic message for the GCA~\cite{kreuk2019hide}, we let the GCA learn the desired message (i.e. adversarial perturbation) and hide it within the input audio file. Finally, we normalize the input of the GCA to zero mean and unit variance to prevent amplitude mismatches between original and adversarial files. 
We save the input's statistics in order to re-normalize the GCA's output, to ensure it has the same mean and standard deviation as the input.

We train the GCA end-to-end, by back-propagating errors captured by two loss functions: a perceptual loss, which accounts for the perceptual differences between the original and adversarial audio files, and an adversarial loss that induces speaker misclassification.   

\subsection{Perceptual and adversarial losses}

It is well known that, in some cases, humans cannot perceive changes introduced to audio files~\cite{kim2000perceptual}.
Motivated by this fact, we design our GCA to learn how to imperceptibly and adversarially perturb an audio file with respect to the human auditory system. To this end, we exploit a very common set of perceptually motivated speech features, called Mel Frequency Cepstral Coefficients (MFCC), that are designed based on the human auditory system ~\cite{sigurdsson2006mel,abdullah2020faults}.
The MFCC computation pipeline is composed of four blocks: Discrete Fourier Transform (DFT) spectrum, Mel filterbank, Log Scaling and Discrete cosine Transform (DCT), which are applied as follows:
1) DFT spectrum:  The audio file is windowed into overlapping frames. Each frame is converted from the time domain to the frequency domain using the Discrete Fourier Transform (DFT);
2) Mel filtering: The DFT spectrum is passed through a set of band-pass filters (called the Mel filterbank), that warp the DFT spectrum to emulate human frequency perception;
3) Log scaling: The log function scales the output of the Mel filterbank to mimic the logarithmic perception of loudness in the human auditory system;
4) DCT: The DCT is then applied on the log-scaled Mel filterbank to decorrelate adjacent frequency bands.

To generate imperceptible adversarial audio files, we define our perceptual loss on MFCC features as follows:
\begin{equation}
    \mathcal{L}_P(\mathbf{x},\dot{\mathbf{x}})=\sum_{t=1}^{T}1-S_{\text{cos}}(\mathbf{f}_t,\dot{\mathbf{f}}_t),
    \label{eq:percept_loss}
\end{equation}
where $S_\text{cos}(\cdot,\cdot)$ is the cosine similarity, defined as:
\begin{equation}
    S_{\text{cos}}(\mathbf{f}_t,\dot{\mathbf{f}}_t)=\frac{\mathbf{f}_t\cdot\dot{\mathbf{f}}_t}{\|\mathbf{f}_t \|\|\dot{\mathbf{f}}_t \|}= \frac{\sum_{i=1}^Ff_i\dot{f}_i}{(\sum_{i=1}^Ff_i^2)^{1/2}(\sum_{i=1}^F\dot{f}_i^2)^{1/2}},
\end{equation}
and $\mathbf{f}_t \in \mathbb{R}^{1\times F}$ and $\dot{\mathbf{f}}_t  \in \mathbb{R}^{1\times F}$ are MFCC feature vectors extracted from the original and perturbed audio files, respectively, at time frame $t$. 
We use the cosine similarity, rather than negated or inverted Euclidean distances, since the latter are sensitive to simple energy differences between the signals, while the former focuses on actual spectral structure.
 
We tailor the learning of the imperceptible perturbation toward misleading the speaker identification model by exploiting an adversarial loss that controls the logits and changes the model's predictions~\cite{carlini2017towards}.
In particular, the untargeted adversarial loss, $\mathcal{L}_{\text{A}_\text{untarg}}(\cdot,\cdot)$, decreases the logits, $\dot{z}_y$ for the actual speaker of the utterance in the adversarial audio file, $y$,
\begin{equation}
\label{eq:Advloss_u}
\mathcal{L}_{\text{A}_\text{untarg}}({\mathbf{x}},\dot{\mathbf{x}})=
\dot{z}_{y} - \max_{{\substack{i=1,...,N\\ i\neq y}}}\dot{z}_i,
\end{equation}
to generate untargeted adversarial audio files that mislead the speaker identification model to output any other speaker:
\begin{equation}
    \dot{y}={f(\dot{\mathbf{x}})\neq y}.
\end{equation}

We also propose a targeted version of our method, FoolHD-t, that aims to force the speaker identification model to predict a specific target class, $y_\text{targ}$, chosen by the attacker:
\begin{equation}
    \dot{y}=y_\text{targ}=f(\dot{\mathbf{x}})\neq y.
\end{equation}
To do this, we use the targeted adversarial loss, $\mathcal{L}_{\text{A}_\text{targ}}(\cdot,\cdot)$, that increases the logit value, $\dot{z}_{y_\text{targ}}$ for the targeted speaker, $y_\text{targ}$, 
\begin{equation}
\label{eq:loss_mislead}
\mathcal{L}_{\text{A}_\text{targ}}({\mathbf{x}},\dot{\mathbf{x}})=
    \max_{{\substack{i=1,...,N \\  i\neq y_\text{targ}}}}\dot{z}_i - \dot{z}_{y_\text{targ}}.
\end{equation}
We select the target speaker randomly~\cite{kurakin2016adversarialscale} among all other speakers.

Our final objective function $\mathcal{L}(\cdot,\cdot)$ accounts for both the perceptual loss, $\mathcal{L}_{\text{P}}(\cdot,\cdot)$, and (untargeted or targeted) adversarial loss, $\mathcal{L}_{\text{A}}(\cdot,\cdot)$, as follows:
\begin{equation}
\label{eq:LossFunction}
    \mathcal{L}({\mathbf{x}},\dot{\mathbf{x}}) = \mathcal{L}_{\text{P}}({\mathbf{x}},\dot{\mathbf{x}}) + \mathcal{L}_{\text{A}}({\mathbf{x}},\dot{\mathbf{x}}).
\end{equation}
The GCA then generates adversarial audio files by minimizing the above objective function over $M$ iterations, with the value of $M$ set experimentally (see Section~\ref{sec:GCA}).

\section{Experimental Setup}
In this section, we describe the dataset used for our experiments, the implementation details of our Gated Convolutional Autoencoder (GCA) and the speaker identification network under attack\footnote{The source code and audio samples are available at \url{https://fsepteixeira.github.io/FoolHD/}}.

\vskip -0.2in
\subsection{VoxCeleb dataset}
The VoxCeleb dataset~\cite{nagrani2017voxceleb} contains speech from 7,363 speakers of multiple ethnicities, accents, occupations and age groups. 
Among these, for our experiments, we randomly chose 250 speakers (for consistency with the state-of-the-art \cite{liZ2020practical,jati2020, xie2020real}): 125 female speakers, and 125 male speakers. All audio files were downsampled to $8$~kHz to match the sampling rate of our pre-trained speaker identification model. Moreover, all files were split into 4 seconds-long segments. Our test set is composed of 10 segments per speaker, amounting to a total of 2,500 segments.

\subsection{Gated Convolutional Autoencoder} 
\label{sec:GCA}
We implemented and trained our GCA in Pytorch using the Adam optimizer with a learning rate of $1\text{e}^{-3}$, similarly to~\cite{kreuk2019hide}. We also used a weight decay of $1\text{e}^{-5}$ and a dropout value of $1\text{e}^{-3}$ to help the network converge. The number of GCA training iterations for generating our adversarial audio files, $M$, was set experimentally as a trade-off between the running time and multi-objective losses. For the untargeted attack, we set $M=500$. However, we increase the value of $M$ to $1,000$ for our targeted attack. While $M=500$ is sufficient to generate imperceptible perturbations that are able to mislead the classifier for the untargeted attack, in the targeted setting the target speaker is potentially far from the original speaker, requiring more iterations to obtain similar results. 
Among the adversarial examples that satisfy our adversarial task (targeted or untargeted), we choose the one with the lowest perceptual loss.

\subsection{Speaker identification model}
\label{sec:spk_id_model}
For our feature extraction stage, we follow Kaldi's \textit{x-vector} recipe \cite{kaldi} and extract 29 MFCC coefficients + log-energy, using 25 ms long windows and 10 ms shift, from each 4 seconds-long file, obtaining 400 frames-long matrices of 
coefficients. Each frame is mean-normalized using a sliding window. Non-speech frames are removed via an energy-based Voice Activity Detection (VAD) module.

We conduct our white-box attack against an \textit{x-vector} speaker identification network, following the architecture of \cite{snyder2018}.
This network is composed of three blocks, a first block that operates at the frame level and two others that operate at the utterance level. The first block is composed of five time-delay layers, with a small temporal context. These layers work as a 1-dimensional convolution, with a kernel size corresponding to the temporal context. In the second block, an attentive statistical pooling layer~\cite{AttentiveStatPool} weighs each time-frame according to its importance and then computes utterance-level statistics 
across the time dimension, providing a summary for the entire speech file. The third and final block takes this summary and propagates it through a set of three fully connected layers, followed by a softmax layer from which a prediction is obtained. All time delay and fully connected layers were followed by a batch normalization layer, a ReLU activation layer and a dropout layer.
This model was implemented in Pytorch and trained over the full dev set of the VoxCeleb dataset (7,323 speakers), for 100 epochs, a learning rate of $1\text{e}^{-3}$, a learning rate decay of $5\text{e}^{-2}$ with a period of 30 epochs, a dropout value of $1\text{e}^{-3}$ and the Adam optimizer. These hyperparameters were optimized for a dev subset consisting of $\sim30\%$ of the dataset. After these parameters were selected, the model was re-trained using all available data. Training samples were augmented with randomly selected Room Impulse Responses (RIRs) and sounds taken from the MUSAN corpus \cite{snyder2018}. This network was further adapted to our 250 test speakers, without data augmentation. To this end, all neurons in the final layer of the network that did not correspond to one of the 250 speakers were dropped. The network was trained for 100 more epochs, using training data from the 250 speakers, and a learning rate of $1\text{e}^{-5}$, with no overlap with our test samples, achieving a final accuracy of 98.3\%.
\section{Results}

We evaluate the performance of FoolHD in terms of effectiveness (i.e. model's accuracy and attack's success rate) and imperceptibility of adversarial audio files for both untargeted and targeted attacks. 
The \textit{untargeted success rate} (S) is defined as the ratio between the number of adversarial audio files that successfully mislead the speaker identification model and the total number of adversarial audio files. The \textit{targeted success rate} (S-t) is defined as the ratio between the number of adversarial audio files that successfully induce the speaker identification model into predicting the target speaker $y_{\text{targ}}$, and the total number of adversarial audio files. Since our speaker identification model is not $100\%$ accurate when classifying our test files, we also report the accuracy 
of the speaker identification model on both the original and adversarial audio files.
To assess the imperceptibility of the adversarial attacks, we use two perceptual audio metrics: Perceptual Evaluation of Speech Quality (PESQ)~\cite{hu2007evaluation} and Just Noticeable Difference (JND)~\cite{Manocha:2020:ADP}. PESQ scores cover a scale from 1 (bad) to 5 (excellent)~\cite{hu2007evaluation}. 
JND is defined as the $l_1$ norm of the difference between the representation of the original and adversarial audio files, computed by a neural network trained using pairs of audio files whose similarity was judged by humans~\cite{Manocha:2020:ADP}.
Our preliminary experiments showed that JND scores are bounded between 0.0 (same audio file) and $\sim5.0$ (signal vs random noise).

\begin{table}[t]
    \centering
    \resizebox{\columnwidth}{!}{
    \setlength\tabcolsep{3pt}
    \begin{tabular}{|l|ccc|cc|}
    \Xhline{3\arrayrulewidth} 
    \multirow{2}{*}{Attack} & \multicolumn{3}{c|}{Effectiveness} & \multicolumn{2}{c|}{Imperceptibility}\\ 
               & Acc.~$\downarrow$    & S~$\uparrow$  &  S-t~$\uparrow$ & JND~$\downarrow$ & PESQ~$\uparrow$ \\
    \Xhline{3\arrayrulewidth} 
    Baseline              &   .983                   &     -       &   -       &     -                 & - \\  
    \hline
    FoolHD-mse            &   .001                   &     .999    &     -     &    $2.93 \pm 1.19$    & $1.44 \pm 0.55$ \\ 
    FoolHD-noSkip         &   .012                   &     .995    &     -     &    $0.97 \pm 0.75$    & $4.34 \pm 0.10$ \\
    FoolHD                &   .012                   &     .996    &     -     &    $0.97 \pm 0.77$    & $4.37 \pm 0.08$ \\\hline 
    FoolHD-t              &   .001                  &     .999    &   .992    &    $1.20 \pm 0.86$    & $4.30 \pm 0.10$ \\

    \Xhline{3\arrayrulewidth} 
    \end{tabular}
    }
    \caption{Impact on effectiveness -- Accuracy (Acc.), Success rate (S) and targeted Success rate (S-t) -- and imperceptibility of using our proposed perceptual loss, skip connection and targeted FoolHD (FoolHD-t).  }
    \label{tab:ablationPerc}
    \vskip -0.2in
\end{table}

\subsection{Ablation study and analysis}

To validate the effectiveness of our perceptual loss and the GCA's skip connection, we present two analyses of FoolHD. 
For the analyses, we consider FoolHD-mse, a modification of the proposed method, where the perceptual loss is replaced by the mean square error (mse) between the input audio file and the adversarial audio file. We further consider another variant of our attack, called FoolHD-noSkip, that has no skip connection from the input to the output of the encoder.
Finally, we also validate the performance of the targeted version of our attack, FoolHD-t.

Table~\ref{tab:ablationPerc} shows the effect of generating adversarial audio files using our proposed perceptual loss and skip connection.
In general, the success rate of adversarial audio files generated by FoolHD, FoolHD-mse and FoolHD-noSkip are similar and above $99\%$. However, the perceptual loss of FoolHD improves the imperceptibility of the adversarial audio files. 
For example, the average PESQ scores of the FoolHD and FoolHD-mse are $4.37$ and $1.44$, respectively.  

When comparing the results of FoolHD and FoolHD-noSkip, we observe that the skip connection provides slight improvements overall. We hypothesize that the skip connection might be preventing vanishing gradients in the backward pass and providing the decoder with information about the original input during the forward pass. Nonetheless, these results are inconclusive and further experimentation is required to assess the skip connection's contribution.

Table~\ref{tab:ablationPerc} also shows that FoolHD-t is $99.2\%$ successful in misleading the speaker identification model into predicting any arbitrary targeted speaker from the original audio file. In addition to this, FoolHD-t drops the accuracy of the speaker identification model to zero.
However, the imperceptibility of FoolHD-t as measured by PESQ and JND is slightly worse than that of FoolHD. 
We hypothesize that this is due to the fact that FoolHD-t needs to introduce more drastic perturbations in the original audio files than FoolHD, in order to not only mislead identification but also reach specific target classes which might be far from the original classes.

\subsection{Other adversarial attacks}
We compare FoolHD with two baseline attacks: the Fast Gradient Sign Method (FGSM) and  the Basic Iterative Method (BIM)~\cite{jati2020adversarial} (we excluded~\cite{wang2020inaudible} from this comparison, as no source code was available at the time of submission). We selected $\epsilon=0.004$ for both methods, to trade-off between effectiveness and imperceptibility of adversarial audio files~\cite{jati2020adversarial}. We use 10 iterations for BIM.

Table~\ref{tab:CompareSOA} compares the effectiveness and imperceptibility of FoolHD with FGSM and BIM.
FGSM achieves only $63.6\%$ success rate in misleading the speaker identification model. However, BIM improves the success rate of FGSM to $100\%$ by iteratively tailoring the adversarial perturbations towards misleading the speaker identification model.
In comparison, the success rate of FoolHD is $99.6\%$. 
While the speaker identification model is still able to recognize correctly  $36.9\%$ of the FGSM adversarial audio files, only $0.4\%$ and $1.2\%$ of BIM's and FoolHD's adversarial audio files are correctly classified by the speaker identification model. 

\begin{table}[t]
    \centering
    \setlength\tabcolsep{3pt}
    \begin{tabular}{|l|cc|cc|}
    \Xhline{3\arrayrulewidth} 
    \multirow{2}{*}{Attack} & \multicolumn{2}{c|}{Effectiveness} & \multicolumn{2}{c|}{Imperceptibility}\\ 
               & Acc.~$\downarrow$    & S~$\uparrow$   & JND~$\downarrow$ & PESQ~$\uparrow$ \\
    \Xhline{3\arrayrulewidth} 
    FGSM~\cite{jati2020adversarial} &   .369                 &     .636        &   $1.29 \pm 1.01$                  &  $3.21\pm0.63$           \\
    BIM~\cite{jati2020adversarial}  &   .004                 &     1.00       &   $1.05 \pm 0.88$   &  $3.36 \pm 0.61$ \\
    \hline
    FoolHD                          &   .012                 &     .996         &    $0.97 \pm 0.77$  & $4.37 \pm 0.08$  \\

    \Xhline{3\arrayrulewidth} 
    \end{tabular}
         \caption{Comparing the effectiveness -- Accuracy (Acc.) and Success rate (S) -- and imperceptibility of FoolHD (untargeted) with Fast Gradient Sign Method (FGSM) and Basic Iterative Method (BIM).}
        \label{tab:CompareSOA}
        \vskip -0.2in
\end{table}

The imperceptibility scores of FGSM and BIM in Table~\ref{tab:CompareSOA} show that bounding the $l_\infty$ norm of adversarial perturbations by $\epsilon$ is not enough for having perceptual similarities between original and adversarial audio files. FoolHD achieves very remarkable improvements in terms of PESQ scores when compared to FGSM and BIM, not only in average but also the standard deviation is significantly lower, showing that there is little variability in the (high) quality of our generated audio files.

\section{Conclusions}

In this paper, we presented a new steganography-inspired method to generate adversarial audio examples. To this end, we trained a Gated Convolutional Autoencoder using a new perceptually motivated multi-objective loss. We showed that our method is capable of generating imperceptible adversarial audio files that are highly successful in attacking a speaker identification model, for both untargeted and targeted scenarios.

The proposed method could be used for several applications, such as privacy protection and watermarking.
As future work, we will test the robustness and transferability of our adversarial audio files to input-based transformations and unseen speaker identification models, respectively. 
Finally, we also aim to extend our method to fool speaker verification systems.

\end{document}